\documentclass[10pt,onecolumn,twoside,english,a4paper]{article}
\usepackage[inner=20mm, right=15mm, top=15mm, bottom=15mm]{geometry}
\usepackage[T1]{fontenc}
\usepackage[latin1]{inputenc}
\usepackage{epsf,epsfig,subfigure,latexsym,latexsym,amssymb,amsmath,alltt}
\usepackage{url}
\usepackage{babel}
\usepackage{balance}
\def\keywords#1{\noindent\textbf{Keywords.} #1}
\def\abstract#1{\noindent\textbf{Abstract.} \textit{#1}}
\def\authorblock[#1]#2{\textbf{\normalsize #1} \\#2}
\def\aff#1{\normalsize #1}

\def\email#1{\texttt{\normalsize #1}}

\newcommand{\A}{A_{ij}}
\newcommand{\B}{B_{ij}}
\newcommand{\C}{C_{ij}}

\newcommand{\E}{E_{ij}}

\newcommand{\BB}{B^{(g)}_{ij}}
\newcommand{\CC}{C^{(g)}_{ij}}

\newcommand{\EE}{E^{(g)}_{ij}}
\newcommand{\DA}{\Delta_A}
\newcommand{\DT}{\Delta_T}
\newcommand{\DDA}{\Delta_A^{(g)}}
\newcommand{\DDT}{\Delta_T^{(g)}}
\newcommand{\R}{R_{ij}}
\newcommand{\RR}{R^{(g)}_{ij}}

\newcommand{\la}{\langle}
\newcommand{\ra}{\rangle}

\title{\textbf{Dynamical networks reconstructed from time series}}
\author{
\authorblock[Zoran Levnaji\'c]{
\aff{Faculty of Information Studies in Novo mesto}\\
\aff{Sevno 13, p.p. 299, 8000 Novo mesto, Slovenia}\\
\email{zoran.levnajic@fis.unm.si}
} }
\date{}

\begin{document}
\maketitle
\pagestyle{empty}
\thispagestyle{empty}

\abstract{Novel method of reconstructing dynamical networks from empirically measured time series is proposed. By examining the variable--derivative correlation of network node pairs, we derive a simple equation that directly yields the adjacency matrix, assuming the intra-network interaction functions to be known. We illustrate the method on a simple example, and discuss the dependence of the reconstruction precision on the properties of time series. Our method is applicable to any network, allowing for reconstruction precision to be maximized, and errors to be estimated.\\}


\section{Introduction}

Complex systems are ubiquitous in nature. On all scales from genes to societies, various systems are composed of many units, which are able to collectively perform complicated tasks despite their simplicity~\cite{mikhailov}. In the recent years, the framework of complex networks was recognized as an excellent formalism for studying complex systems. By representing units as nodes and modeling their interactions as links~\cite{dorogo}, science of complex networks introduced graph analysis methods into physics, biology, engineering and even sociology~\cite{costa}. This allowed for a variety of real and artificial complex systems to be extensively examined, typically via computational modeling~\cite{us-pre1}. Crucial aspect of a complex network is its structure, i.e. the topology of connections among its nodes. Properties of network structure dictate its global behavior, and are key to understanding the network's functioning and potentials for its control. For simple oscillator models, profound intertwinement between network structure and network dynamics was recently shown~\cite{me}.

Since the structure of many natural networks is only partially known, it is of central interest to develop methods for reconstructing the network structure from the available empirical information. Various experimental techniques in this directions are already in use, specially in the context of gene regulation networks~\cite{geier-hecker}. In addition, a range of mathematical results is available~\cite{lu}. Recently, the topology of a social network was inferred using mobile phone data~\cite{eagle}. \textit{Invasive} reconstruction methods involve perturbing the network dynamics which allows for structural data to be easily extracted~\cite{us-timme}. Although invasive methods generally give good results, it is often unpractical to interact with the on-going network dynamics. \textit{Non-invasive} reconstruction methods focus on investigation of the observable network outputs, such as the time series quantifying the system's dynamics~\cite{shandilya,kralemann,hempel,xia}. The relevance of non-invasive approach is increasingly recognized, particularly due its suitability for detecting links in biological networks~\cite{hempel,xia}. Alternatively, reconstruction methods also rely on techniques from control theory~\cite{dongchuan}, and even compressive sensing~\cite{wang}.

In this contribution, we propose a novel network reconstruction method, based on examining the correlations between the variables and the derivatives corresponding to different nodes. Our central assumption is the precise knowledge of the functional forms of the intra-network interactions. As we show, depending on the quantity of network information contained in the empirical data, our method can give very precise results even for very short time series, thus being suitable for actual experimental situations. Apart from being non-invasive, our method is conceptually very simple, and easy to numerically implement. In contrast to a recent result based on the same hypothesis~\cite{shandilya}, our method avoids solving the overdetermined linear system, and allows for the reconstruction error to be estimated.


\section{The Reconstruction Model}

We consider a complex system composed of $N$ interacting units, which we represent as a network with $N$ nodes, whose links model the interaction between the nodes. Each node is assigned a dynamical state defined by the real variable $x_i \equiv x_i (t)$, where $i = 1,\hdots N$. We assume to be in the possession of empirically obtained discrete-time trajectories $x_i (t_m)$ which describe the system's dynamical evolution over a certain time interval. The available data consists of $N$ sequences, each containing $L$ values $x_i(t_1),\hdots x_i(t_L)$. The measurements of $x_i$ are separated by the observation interval $\delta_t = t_{m+1}-t_m$, which defines the resolution of the time series (sampling frequency). Time interval $\delta_t$ is uniform and assumed smaller than the characteristic dynamical time scale.

We further assume the time-evolution of the node $i$ to be given by: 
\begin{equation} \dot x_i = f_L (x_i) + \sum_{j=1}^N A_{ji} f_C (x_j)  \, , \label{eq-1} \end{equation}
where we describe the local dynamics via function $f_L (x)$, and the network (inter-node) interaction by the function $f_C$. The network structure is encoded in the adjacency matrix $\A$, whose element $ij$ specifies the strength with which the node $i$ acts on the node $j$. The dynamics of the node $i$ is a cumulative effect of its local dynamics and the sum of contributions from its networks neighbors that come with different strengths. Finally, we also assume that both interaction functions $f_L$ and $f_C$ are precisely known.

We seek to reconstruct the network's adjacency matrix $\A$ under the named assumptions -- by having the ``fingerprint'' of system's behavior, we attempt to reveal its structure. The above assumptions on which we build our theory are realistic. Many natural systems are modeled using Eq.\ref{eq-1}: examples include gene regulation and neural interactions, for which the interaction functions are widely investigated, and do not vary with network links. Modern experimental techniques allow for high-resolution measurements of quantities such as gene expression, although thus obtained time series are typically short.

When examining inter-dependence between dynamical quantities, one is typically interested in calculating the correlation between two dynamical variables. Inspired by this, we construct our theory based on investigating the correlation between a variable ($x_i$) and the derivative of another variable ($\dot x_j$). We hence examine the correlation between the motion of the node $i$, and the speed of node $j$. We start by defining the following matrices:
\begin{equation} \begin{array}{lll}
  B_{ij} &=& \langle x_i \dot x_j \rangle  \; ,  \\ 
  C_{ij} &=& \langle x_i f_L (x_j) \rangle  \; , \\  
  E_{ij} &=& \langle x_i f_C (x_j) \rangle  \; ,       \label{eq-2}
\end{array} \end{equation}
where $\la \cdot \ra$ denotes the time-average of a dynamical quantity (i.e., average over the recorded time-evolution) $\la h \ra = \frac{1}{L} \sum_{m=1}^L h (t_m)$. This allows for Eq.\ref{eq-1} to be re-written in the matrix form:
\begin{equation}  \A = ( B_{ik} - C_{ik} ) \cdot E_{kj}^{-1} \; , \label{eq-3} \end{equation}
which is our main network reconstruction equation. We introduce a new set of time points:
\[ \tau_m = \frac{t_{m+1} - t_m}{2} \; , \;\; m = 1 , \hdots, L-1 \; , \]
so that 
\[  \dot x_i (\tau_m) =  \frac{x_i (t_{m+1}) - x_i (t_m)}{\delta_t} \; ,\]
and accordingly:
\[  f_{L,C} (\tau_m) =  \frac{f_{L,C} (t_{m+1}) + f_{L,C} (t_m)}{2} \; . \]
This provides a more stable estimation of both interaction function values and the derivative values. We will rely on this calculation scheme for the implementation of our theory through Eq.\ref{eq-3}. Note that in principle, our method is applicable to any network, and the reconstruction is precisely correct in the limit of very long time series. However, since the empirical data are not only finite, but typically very short, our method will in general yield an approximate reconstruction.

To discern from the original adjacency matrix $\A$, we term the reconstructed adjacency matrix $\R$, and quantify the matrix reconstruction error as follows:
\[ \DA = \sqrt{  \frac{   \sum_{ij} [ \R - A_{ij} ]^2}{ \sum_{ij}  A_{ij}^2 } } \, . \]
Natural test to make for each obtained $\R$ is to quantify how well does it reproduce the original empirical data $x_i (t_m)$. To achieve this, we apply the following procedure: start the run from $x_i (t_1)$ for all nodes, and run the dynamics using adjacency matrix $\R$ for the time interval $\delta_t$, i.e. until the time $t_2$. Denote thus obtained values $y_i (t_2)$, re-start the run from $x_i (t_2)$ running until $t_3$, accordingly obtaining $y_i (t_3)$, and so on. The discrepancy that the time series $y_i (t_m)$ show in comparison to $x_i (t_m)$ is the most straightforward measure of the reconstruction precision for matrix $\R$. We name it trajectory error $\DT$, and define as follows:
\[  \DT = \frac{1}{N} \sum_i  \sqrt{ \frac{ \sum_m [ x_i (t_m) - y_i (t_m) ]^2 }{  \sum_m [ x_i (t_m) - \bar{x}_i ]^2 }  } \; . \]
This way we measure point-by-point exactness of the reconstructed trajectory, which quantifies how well does it conform to the empirical data. Small $\DT$ is necessary, but not sufficient for a good reconstruction -- easily reproducible time series (such as periodic orbits) always display very small $\DT$ regardless of $\DA$, since many different networks can produce such data. On the other hand, hardly reproducible time series (such as transient or chaotic orbits) may show large $\DT$ that does not imply large $\DA$. However, as we show, two errors are in general related, allowing for estimation of $\DA$ based on $\R$ and $\DT$ only.


\section{Results}

\begin{figure}[!hbt] \centering 
\includegraphics[width=0.4\textwidth]{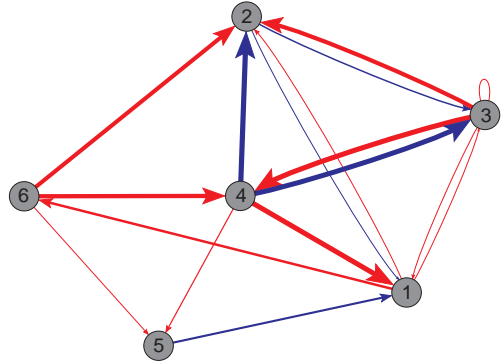} \caption{Graphical representation of the studied network. Link thickness illustrates the interaction strength. Red/blue link colors (light/dark shades) indicate positive/negative inter-node interactions. Nodes are numbered in accordance with the adjacency matrix shown in Fig.\ref{figure-3}a.} 
\label{figure-1} \end{figure}

\begin{figure}[hbt!] \centering 
\includegraphics[width=0.91\textwidth]{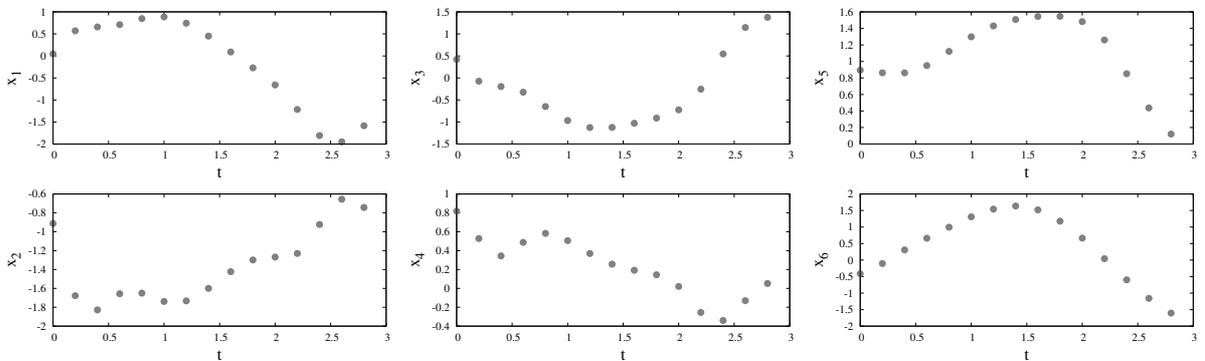} \caption{Time series for all 6 nodes for network Fig.\ref{figure-1}, obtained for the first set of initial conditions.} 
\label{figure-2}  \end{figure}

\begin{figure}[!hbt]  \centering  
\includegraphics[width=0.91\textwidth]{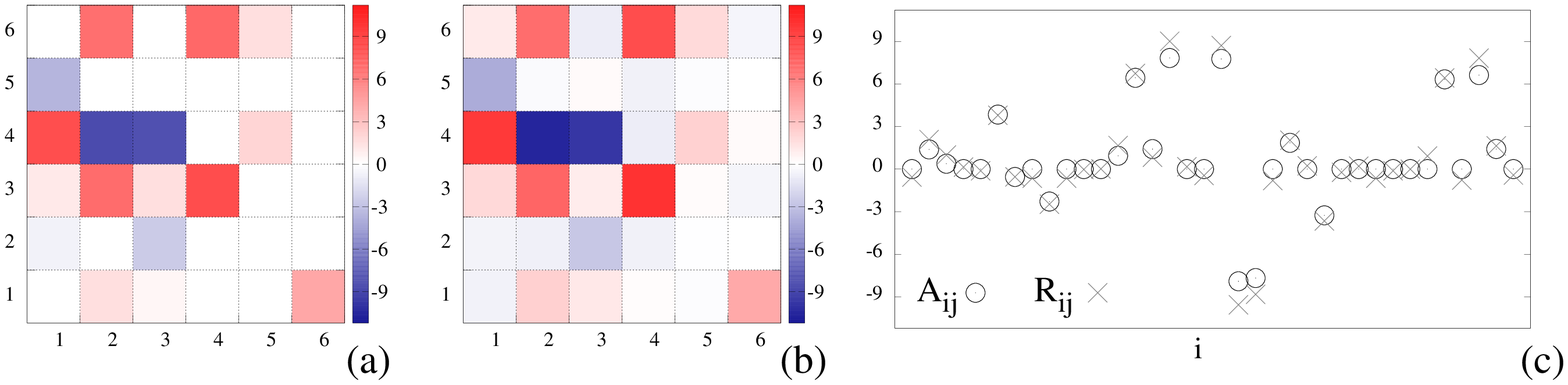}
\caption{Original $\A$ and reconstructed $\R$ adjacency matrices in (a) and (b) respectively. Colorbar (shade) indicates the weights obtained from time series Fig.\ref{figure-2}. Link-by-link weights comparison of $\A$ (circles) and $\R$ (crosses) in (c). Matrix error $\DA=0.18$, trajectory error $\DT=0.038$.} 
\label{figure-3} \end{figure}

We test our reconstruction method using a simple illustrative example. A network with $N=6$ nodes is constructed by placing $L=17$ directed links between randomly chosen pairs of nodes, while requiring the resulting network to be connected. Links are weighted with positive and negative weights, uniformly selected at random from $[-10,10]$. The studied network is illustrated in Fig.\ref{figure-1}, and its adjacency matrix $\A$ is shown in Fig.\ref{figure-3}a.
The dynamics is defined on the network via Hansel-Sompolinsky model~\cite{HS} by putting $f_L = -x$ and $f_C = \tanh x$ in Eq.\ref{eq-1}. The complete dynamics on network reads: 
\begin{equation} \dot x_i = - x_i + \sum_{j=1}^{6}  A_{ji} \tanh (x_j)  \; . \label{eq-4} \end{equation}
For each node we randomly select an initial condition from $[-1,1]$, and numerically integrate Eq.\ref{eq-4} from time $t=0$ to $t=3$. During the run, we store 15 values for each $x_i$, equally spaced in time, starting with $x_i (t_1=0)$. Thus obtained time series for all nodes are shown in Fig.\ref{figure-2}. 

We assume now these time series to be obtained from an ``external'' source (e.g. coming from experimental measurement), and seek to employ them to reconstruct the network's adjacency matrix as discussed in the previous Section. To this end, we numerically compute the matrices $\B$, $\C$ and $\E$, and obtain the reconstructed adjacency matrix $\R$ via Eq.\ref{eq-3}. The result is shown in Fig.\ref{figure-3}  -- the original $\A$ in (a), the reconstructed $\R$ in (b), and link-by-link comparison of $\A$ and $\R$ in
(c). The reconstructed matrix $\R$ reasonably well approximates the original $\A$, both for zero and non-zero weights. The matrix error is $\DA = 0.18$, and the trajectory error is $\DT = 0.038$, indicating a good reconstruction precision.

\begin{figure}[!hbt] \centering 
\includegraphics[width=0.91\textwidth]{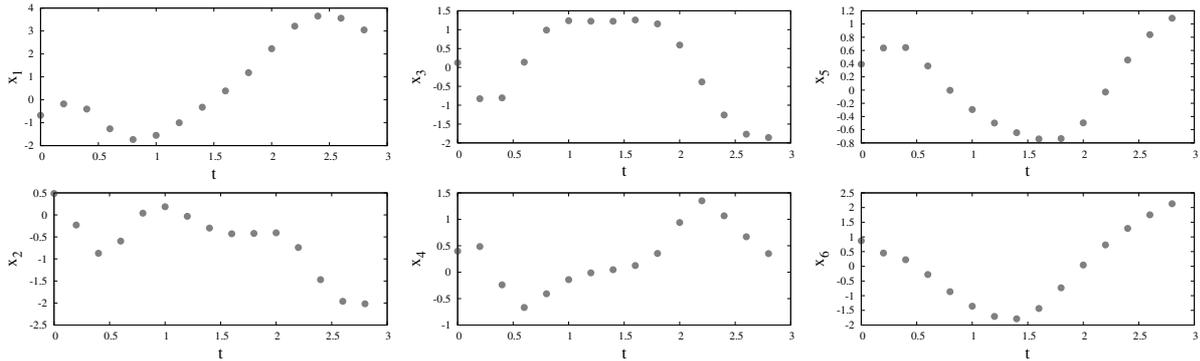}
\caption{Time series for all 6 nodes for network Fig.\ref{figure-1}, obtained for the second set of initial conditions.} 
\label{figure-4} \end{figure}

\begin{figure}[!hbt]  \centering  
\includegraphics[width=0.91\textwidth]{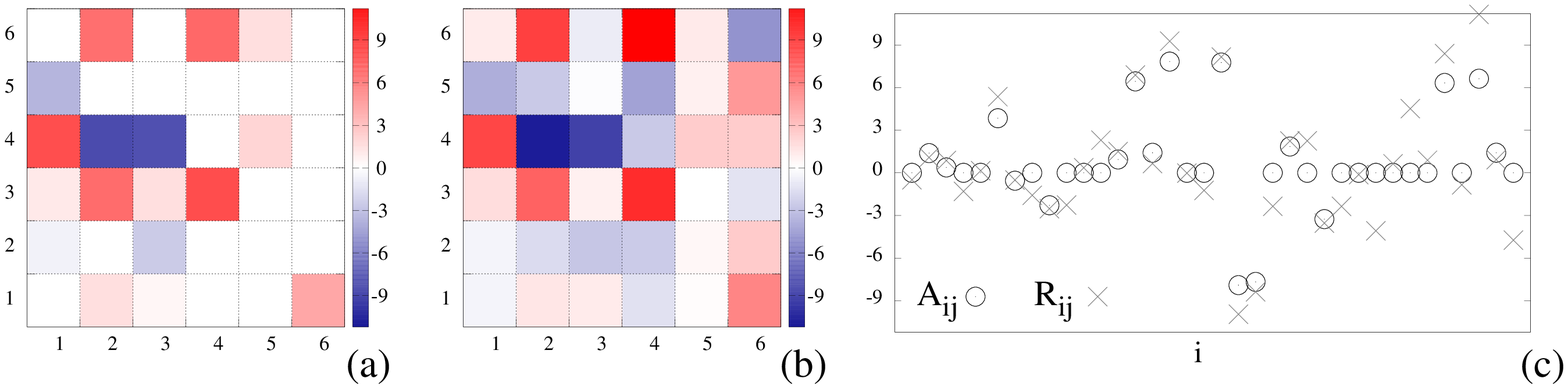}
\caption{Original $\A$ and reconstructed $\R$ adjacency matrices in (a) and (b) respectively. Colorbar (shade) indicates the weights obtained from time series Fig.\ref{figure-4}. Link-by-link weights comparison of $\A$ (circles) and $\R$ (crosses) in (c). Matrix error $\DA=0.56$, trajectory error $\DT=0.05$.} 
\label{figure-5} \end{figure}

We now run another simulation of our dynamical system Eq.\ref{eq-4} with the same underlying network, but this time starting from a different set of initial conditions. A new set of time series of equal size and resolution is obtained and shown in Fig.\ref{figure-4}, from which we seek to reconstruct our network again.
The new results are shown in Fig.\ref{figure-5}, and organized in analogy with Fig.\ref{figure-3}.
The new $\R$ has matrix error $\DA=0.56$ and a trajectory error of $\DT=0.05$, which is considerably worse than in the previous example, as it can also be clearly seen by comparing Fig.\ref{figure-3}c and Fig.\ref{figure-5}c.

Despite that both sets of time series were produced by the same dynamical network, two reconstructed networks are different. This shows that besides depending on the length and resolution of the time series, the reconstruction precision crucially depends on the ``quality'' of time series as well, i.e. on the quantity of network information contained in them. Easily reproducible data contains less information than hardly reproducible data. As just illustrated, a given dynamical network can yield different time series depending on the initial conditions. Is there a relation between the two errors that could be used to estimate $\DA$ based only on $\DT$, independently on the ``quality'' of time series? The final Section of this paper is devoted to providing at least a  preliminary answer to this question.


\section{Discussion} 

The proposed reconstruction method in principle applies to any network whose inter-node interactions can be described via Eq.\ref{eq-1}. The final reconstruction precision depends on a number of factors: (\textit{i}) length and resolution of time series, also related to the precision of derivative estimates; (\textit{ii}) quantity of network information contained in the empirical data, which can be seen as reproducibility of the time series, or coverage of the dynamical phase space with data; (\textit{iii}) invertibility of the matrix $\E$; and finally, (\textit{iv}) properties of the network itself -- some networks can be more reconstructable than others. In a concrete reconstruction problem, it is difficult to isolate how much each factor contributes to $\DA$. Instead of quantifying this, we show a generalization of our method, done towards improving the reconstruction precision and estimating reconstruction errors.

Our method is based on calculating the correlations between the variable $x_i$ and other terms, as defined in Eq.\ref{eq-2}. More generally, we can replace $x_i$ by $g(x_i)$, where $g$ is an arbitrary function, without changing the main result. Eq.\ref{eq-2} now becomes:
\begin{equation} \begin{array}{lll}
  \BB &=& \langle g(x_i) \dot x_j \rangle  \; ,  \\ 
  \CC &=& \langle g(x_i) f_L (x_j) \rangle  \; , \\  
  \EE &=& \langle g(x_i) f_C (x_j) \rangle  \; ,       \label{eq-5}
\end{array} \end{equation}
where notation $\BB$ indicates that the matrix $\B$ was calculated via Eq.\ref{eq-5} using function $g$. Eq.\ref{eq-3}, which now reads:
\begin{equation}  \RR = \big( \BB - \CC \big) \cdot \EE\;^{-1} \; , \label{eq-6} \end{equation}
still holds for any function $g$. As before, in the limit of very long time series, the reconstruction is precisely correct for any choice of $g$. For realistic scenarios involving very short time series, the reconstruction precision will depend on $g$, as two different $g$-s will in general yield two different $\RR$-s. This means that $g$ plays the role of a tunable parameter, which can be used to find the best reconstruction. By considering a set of functions $g$, we can compute $\RR$ for each of them, and define as the best reconstruction that $\RR$ whose reconstructed dynamics shows minimal $\DDT$ (we generalize matrix and trajectory errors to depend on $g$). A good choice of $g$ will extract more extractable network information hidden in the empirical data, and improve the simple reconstruction for $g(x)=x$. Moreover, variations of $\RR$ with $g$ are related to the reconstruction precision -- for a reliable reconstruction, the obtained $\RR$ will not strongly depend on changes of $g$. On the other hand, a bad reconstruction will be recognized by a drastic dependence of $\RR$ on $g$. Note that the functional properties of $g$ itself are irrelevant -- the \textit{only} role of $g$ is the computation of $\RR$.

\begin{figure}[!hbt]  \centering  
\includegraphics[width=0.5\textwidth]{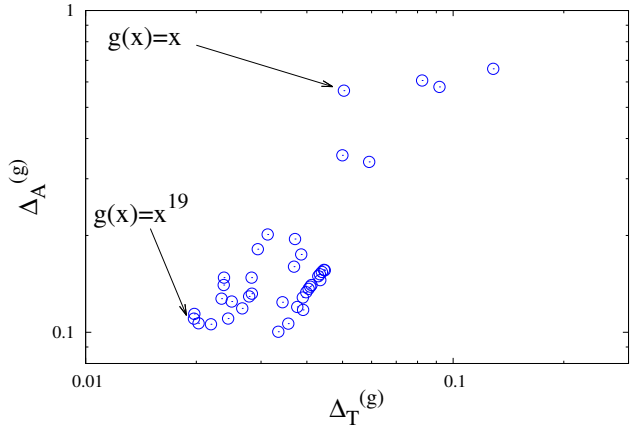}
\caption{Reconstructions using many functions $g$. Each point is specified by $\DDT$ and $\DDA$ corresponding to $\RR$ obtained via one of $g(x) = x^n$ for $n$ integer between -20 and 20 except 0. Points obtained for $g(x)=x$ and $g(x)=x^{19}$ are indicated.} 
\label{figure-6} \end{figure}

\begin{figure}[!hbt]  \centering  
\includegraphics[width=0.91\textwidth]{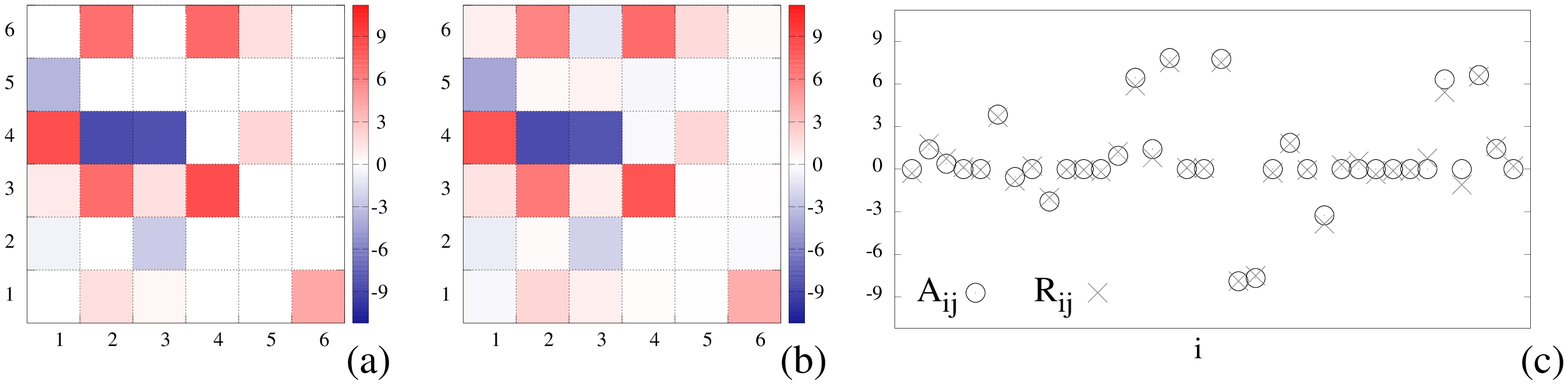}
\caption{Original $\A$ and reconstructed $\RR$ adjacency matrices in (a) and (b) respectively. Colorbar (shade) indicates the weights obtained from time series Fig.\ref{figure-4}, using the function $g(x)=x^{19}$. Link-by-link weights comparison of $\A$ (circles) and $\RR$ (crosses). Matrix error $\DDA=0.11$, trajectory error $\DDT=0.02$.} 
\label{figure-7} \end{figure}

To illustrate the implementation of our generalized method, we examine again the second set of time series shown in Fig.\ref{figure-4}. We consider the set of functions $g(x) = x^n$, where for $n$ we take integers between -20 and 20 (except 0). The network is reconstructed using Eq.\ref{eq-6} for each $g$, and the corresponding $\DDA$ and $\DDT$ are calculated. The results are shown in Fig.\ref{figure-6}, where each $\RR$ is represented through its $\DDT$ and $\DDA$. 
There is a visible correlation between the two errors, suggesting that smaller $\DDT$, on average, leads to a smaller $\DDA$. Following this principle, for function $g(x)=x^{19}$ we find the smallest trajectory error $\DDT=0.02$ leading to $\DDA=0.11$. As indicated in Fig.\ref{figure-6}, this result is much better than what obtained for $g(x)=x$. In addition, this result is better than the one found for time series from Fig.\ref{figure-2}. For comparison, we show the reconstruction for $g(x)=x^{19}$ in Fig.\ref{figure-7}.
Note however, that this is not the best result in terms of $\DDA$: for $g(x)=x^{-6}$ we find $\DDA=0.10$, that however we missed since it has bigger trajectory error $\DDT=0.033$. This indicates that considering few $g$-s with the smallest $\DDT$-s we can construct an error bar on each element of $\RR$, thus defining a confidence interval for each reconstructed value of $\RR$. As clear from Fig.\ref{figure-6}, a cluster of points around $g(x)=x^{19}$ is a good candidate for such set of $\RR$-s. Of course, considering many linearly independent $g$-s from a given functional family would yield a much better best $\RR$ and more confident error bars.

These findings suggest that through the appropriate tuning of $g$, we can compensate for the ``low quality'' of time series, which can considerably improve the reconstruction precision, and even allow for estimation of $\DA$. The question of selecting the optimal function $g$ which extracts all the network information contained in the time series remains open. Most straightforwardly, one can search for such $g$ via Monte Carlo method using many linearly independent functions. An intriguing result would be a way to design the optimal $g$ based on the time series and interaction functions. On the other hand, the best $g$ might be obtainable through techniques such as evolutionary optimization algorithms or machine learning.

We finish the paper by discussing the limits and proposing further extensions of our method. Our strongest hypothesis is the precise knowledge of interaction functions. Despite the availability of good mathematical models for many natural interactions, lifting this assumption would greatly enhance the generality of our theory. When approximate functional forms are known, interaction functions can be expanded in series, facilitating their reconstruction. This would mean that for each $g$, we obtain not just $\RR$, but also $f_L^{(g)}$ and $f_C^{(g)}$. This leads to a possibility of obtaining many different networks, all reproducing empirical data equally well, but in pair with different interaction functions. Another extension regards our assumption that the mathematical form of interactions is given by Eq.\ref{eq-1}. While a similar theory could be developed for any known form of Eq.\ref{eq-1}, the problem arises for networks whose interactions form is not known. Furthermore, since noise is present in all physical processes and experimental measurements, our method should be applicable to noisy empirical data. Finally, we note that our problem of network reconstruction is similar to the problem of designing a network with prescribed dynamics. One can use our method to design a network that displays given time series, by specifying the tolerance in $\DT$.


\section{Acknowledgments}

Many thanks to Arkady for contributing the original idea, in addition to Misha, \v Suki and Bernard for constructive suggestions.


\begin{thebibliography}{}
\bibitem{mikhailov} A.~S.~Mikhailov; V.~Calenbuhr. \textit{From Cells to Societies: Models of Complex Coherent Action}. Springer, 2006.
\bibitem{dorogo} S.~N.~Dorogovtsev. \textit{Lectures on Complex Networks}. Oxford University Press, 2010.
\bibitem{costa} L.~F.~Costa \textit{et al}. Analyzing and modeling real-world phenomena with complex networks: a survey of applications. \textit{Adv. Phys.} 60(3), 329, 2011.
\bibitem{us-pre1} Z.~Levnaji\'c; A.~Pikovsky. Phase resetting of collective rhythm in ensembles of oscillators. \textit{Phys. Rev. E} 82, 056202, 2010.
\bibitem{me} Z.~Levnaji\'c. Emergent multistability and frustration in phase-repulsive networks of oscillators. \textit{Phys. Rev. E} 84, 016231, 2011. 
\bibitem{geier-hecker} F.~Geier; J.~Timmer; C.~Fleck. Reconstructing gene-regulatory networks from time series, knock-out data, and prior knowledge. \textit{BMC Sys. Biol.} 1, 11, 2007. M.~Hecker \textit{et al.} Gene regulatory network inference: Data integration in dynamic models - A review. \textit{Biosystems} 96(1), 86, 2009.
\bibitem{lu} L.~L\"u; T.~Zhou. Link prediction in complex networks: A survey. \textit{Physica A} 390(6), 1150, 2011.
\bibitem{eagle} N.~Eagle; A.~Pentland; D.~Lazer. Inferring friendship network structure by using mobile phone data. \textit{Proc. Nat. Acad. Sci. USA} 106(36), 15274, 2009.
\bibitem{us-timme} Z.~Levnaji\'c; A.~Pikovsky. Network Reconstruction from Random Phase Resetting. \textit{Phys. Rev. Lett.} 107, 034101, 2011. M.~Timme. Revealing Network Connectivity from Response Dynamics. \textit{Phys. Rev. Lett.} 98, 224101, 2007.
\bibitem{shandilya} S.~G.~Shandilya; M.~Timme. Inferring network topology from complex dynamics. \textit{New J. Phys.} 13, 013004, 2011.
\bibitem{kralemann} B.~Kralemann; A.~Pikovsky; M.~Rosenblum. Reconstructing phase dynamics of oscillator networks. \textit{CHAOS} 21, 025104, 2011.
\bibitem{hempel} S.~Hempel \textit{et al.} Inner Composition Alignment for Inferring Directed Networks from Short Time Series. \textit{Phys. Rev. Lett.} 107, 054101, 2011.
\bibitem{xia} Q.~Xia \textit{et al}. Inference of gene regulatory networks with the strong-inhibition Boolean model. \textit{New J. Phys.} 13, 083002, 2011.
\bibitem{dongchuan} D.~Yu; M.~Righero; Lj.~Kocarev. Estimating Topology of Networks. \textit{Phys. Rev. Lett.} 97,  188701, 2006.
\bibitem{wang} W.~Wang. Network Reconstruction Based on Evolutionary-Game Data via Compressive Sensing. \textit{Phys. Rev. X} 1, 021021, 2011.
\bibitem{HS} D.~Hansel; H.~Sompolinsky. Solvable Model of Spatiotemporal Chaos. \textit{Phys. Rev. Lett.} 71, 2710, 1993.
\end{thebibliography}
\end{document}